\documentstyle[11pt]{article}
\begin{document}
\title{Probing the Isospin Dependent In-Medium Nucleon-Nucleon 
Cross Section by Nucleon Emissions}
\bigskip
\author{\bf .Jian-Ye Liu$^{1,2}$, Wen-Jun Guo$^{2}$,Yong-Zhong Xing$^{2,4}$\\
\bf Wei Zuo$^{1,2}$, Xi-Guo Lee$^{1,2}$, Zeng-Hua Li$^{2}$}

\date{}
\maketitle
\begin{center}
$^{1}$Center of Theoretical Nuclear Physics, National Laboratory of
Heavy Ion Accelerator,\\ Lanzhou 730000, P.R.China\\
$^{2}$Institute of Modern Physics, Chinese Academy of Sciences,\\ 
Lanzhou 730000, P.R.China\\
$^{4}${\small Department of Physics, Tianshui Normal College, Gansu Tianshui, 741000,P.R.China}\\
\end{center}
\maketitle
\begin{center}{\bf Abstract}
\hskip 0.3in .cThe effects of the symmetry potential and the isospin dependent in-medium
nucleon-nucleon (NN) cross section on the number of proton(neutron) emissions N$_p$(N$_n$)
are studied respectively within an isospin-dependent quantum molecular dynamics (IQMD) model.
The isospin dependent in-medium NN cross section is found to have a strong influence
on N$_p$(N$_n$) but N$_p$(N$_n$) is not sensitive to the symmetry potential for the 
neutron-deficient colliding system at relatively high energies.We propose to make use 
of the N$_p$(N$_n$) as a probe to extract information on the isospin dependent 
in-medium NN cross section.

\par
{\bf PACs Number(s)}: 25$\cdot$70$\cdot$pg,02$\cdot$70$\cdot$ $Ns$,24$\cdot$10$\cdot$Lx\\
{\bf Keywords}: isospin effect,nucleon emission,nucleon-nucleon cross section,
symmetry potential
\end{center}
\par
\par
\begin{center}
\section{Introduction}
\hskip 0.3in With the rapid advance in radioactive beam physics a better 
understanding of the isospin degree of freedom in nuclear collision dynamics
may provide us with useful hints on how to extract reliable information about
the in-medium NN cross section and the symmetry potential [1-5].
To obtain this information several interesting isospin effects in heavy ion 
collisions have been explored both experimentally [6-16] and theoretically,
e.g.,[17-25] over the last few years.Bao-An Li et al.[19,25] have found that
information on the symmetry potential can be extracted by studying
the neutron to proton ratio of preeqilibrium emissions in the relative low
energy heavy ion collisions (HIC).In contrast,in this paper our studies within
IQMD found that the numbers of proton(neutron) emissions N$_p$(N$_n$) depend 
sensitively on the isospin dependent in-medium NN cross section and weakly
on the symmetry potential in the relatively high beam energy region for the 
neutron-deficient systems.In this case,N$_p$(N$_n$) is a probe for extracting
 information about the isospin dependence of in-medium NN cross section
in the relative high beam energy region for the neutron-deficient systems.
\par
\end{center}
\section{Theoretical Model}
\hskip 0.3in 
The dynamics of intermediate energy heavy ion collisions
described by QMD contains two ingredients: density dependent
mean field and in-medium N-N cross section.To describe
isospin effects appropriately, QMD should be modified properly: the density
dependent mean field should contain the correct isospin terms including
symmetry potential and Coulomb potential,the in-medium N-N cross section
should be different for neutron-neutron ( proton-proton ) and neutron-proton
collisions in which Pauli blocking should be counted by distinguishing neutrons from
protons.In addition,the initial condition of the ground state of two
colliding nuclei should also contain isospin information.The main physics
ingredients and their numerical realization in the IQMD model can be found in Refs.
[3,28,30,31,34].In the IQMD model, the density distributions of colliding nuclei
were from the calculations of the Skyrme- Hatree-Fock with parameter set
SKM$^{*}$ $^{[32]}$ and the initial code of IQMD was used to determine the
ground state properties of the colliding nuclei,such as the binding energies
and rms radii which are the same as the experimental data.In the presence
calculations the interaction potential  in the IQMD were determined 
as follows:
\begin{equation}
U(\rho )=U^{Sky}+V^{Coul}+U^{sym}+V^{Yuk}+U^{MDI}+U^{Pauli}
\end{equation}
$U^{Sky}$ is the density-dependent Skyrme potential,
\begin{equation}
U^{Sky}=\alpha (\frac \rho {\rho _0})+\beta (\frac \rho {\rho _0})^\gamma
\end{equation}
$V_c$ is Coulomb potential. $U^{Yuk}$ is the Yukawa potential ,
\begin{equation}
U^{Yuk}=t_3exp(\frac{\left|\overrightarrow{r_1}-
\overrightarrow{r_2}\right|}{m})/\frac{\left|\overrightarrow{r_1}-
\overrightarrow{r_2}\right|}{m}
\end{equation}
 $U^{MDI}$ is the momentum dependent interaction,
\begin{equation}
U^{MDI}=t_4ln^2[t_5(\overrightarrow{p_1}-\overrightarrow{p_2})^2+1]\frac
\rho {\rho _0}
\end{equation}
$U^{Pauli}$ is the Pauli potential ,
\begin{equation}
U^{Pauli}=V_p\{\frac \hbar {p_0q_0})^3exp(-\frac{(\overrightarrow{r_i}-%
\overrightarrow{r_j})^2}{2q_0^2}-\frac{(\overrightarrow{p_i}-\overrightarrow{%
p_j})^2}{2p_0^2}\}\delta _{p_ip_j}
\end{equation}
\[
\delta _{p_ip_j}=\left\{
\begin{array}{ll}
1 & \mbox{for neutron-neutron or proton-proton} \\
0 & \mbox{for
neutron-proton}
\end{array}
\right.
\]
$U^{sym}$ is the symmetry potential.In the present calculation, two
different density dependent symmetry potentials $^{[2,3,29]}$ are used,
i.e., $U_1^{sym}=cF_1(u)\delta \tau _z$ and $U_2^{sym}=cF_2(u)[\delta \tau _z+%
\frac{1}{2}\delta ^2]$, where $\tau_{z}=1$ for neutron and $\tau_{z}=-1$ for
proton, $F_1(u)=u$ and  $F_2(u)=u^2$ , $u\equiv \rho/
\rho_{0} $. $\delta$ is the relative neutron excess $\delta =\frac{%
\rho_n-\rho_p}{\rho_n+\rho_p} =\frac{\rho_n-\rho_p}{\rho} $.Here c is the
strength of symmetry potential, taking the value of 32 or 0 MeV (the c=0.0 case
is denoted by $U_0^{sym}$). $\rho$,$\rho _{_0}$, $\rho _n$ and
$\rho _p$ are total density and its normal value, neutron density and proton density,
respectively. It is worth mentioning that the recent studies on collective flow
in HIC at intermediate energies have indicated a reduction of in-medium N-N
cross sections. An empirical expression of the density dependent in-medium N-N
cross section$^{[26]}$ is given by
\begin{equation}
\sigma^{med}_{NN}=(1+\alpha\frac{\rho}{\rho_{0}})\sigma^{free}_{NN}
\end{equation}
Where the parameter $\alpha$ $\approx-0.2$ has been found to reproduce
the flow data. $\sigma^{free}_{NN}$ is the experimental
N-N cross section$^{[27]}$.
The parameters of the interaction potentials
are given in table 1.

\begin{center}

\scriptsize

\begin{tabular}{|c|c|c|c|c|c|c|c|c|c|}\hline
$\alpha$ & $\beta$ &$\gamma$&$t_{3}$&m&$t_{4}$&$t_{5}$&$V_{P}$&$p_{0}
$&$q_{0}$\\ \hline
(MeV)&(MeV)&&(MeV)&(fm)&(MeV)&($MeV^{-2}$)&(MeV)&(MeV/c)&(fm)\\\hline
-390.1&320.3&1.14&7.5&0.8&1.57&$5\times10^{-4}$&30&400&5.64\\\hline
\end{tabular}\\
\vskip 0.3in
Table 1.The parameters of the interaction potentials
\end{center}
The free neutron-proton cross section $\sigma^{free}_{NN}$ is about a
factor of 3 times larger than the free proton-proton or free neutron-neutron cross section
below about 400 MeV (in the Lab).It is worth to mentioning that the relationship between the neutron-proton
cross section and neutron-neutron (proton-proton) cross section depends also on the modification
of the nuclear density distributions during reactions.
We construct clusters by means of the isospin-dependent modified coalescence model
$^{[34]}$,in which particles with relative momentum smaller than $p_{0}$= 300MeV/c
and relative distance smaller than $R_{0}$= 3.5 fm are used.We make use of
the restructured aggregation model$^{[35]}$ to avoid the nonphysical clusters
after constructing the clusters, until there are not any nonphysical clusters
to be produced.
\section{Results and Discussions}
\hskip 0.1in 
\subsection{Checking the IQMD model }
In order to check the IQMD code with the above parameters,
the multiplicity of the intermediate mass fragments
$N_{imf}$ for the reactions $^{58}Fe+^{58}Fe$ and
$^{58}Ni+^{58}Ni$ at the beam energy $E =75$MeV/u has been calculated
by using the IQMD code with the above parameters.The multiplicity of the intermediate
mass fragments (IMFs) is defined as the number of fragments with charge numbers
from 3 to 18.The calculated results are compared with the experimental data $^{[36]}$
on the same scale in Fig.1 which gives the correlation
between the mean value of the intermediate mass fragment multiplicity
$N_{imf}$ and the charged particle multiplicity $N_{c}$. The
solid (open) circles represent the experimental data for the
reaction $^{58}Ni+^{58}Ni$ ($^{58}Fe+^{58}Fe$) at $E=75$MeV/u and
the solid line (dot line) denotes the IQMD results for
$^{58}Ni+^{58}Ni$ ($^{58}Fe+^{58}Fe$).It is clear that the
present IQMD predictions are in satisfactory agreement with
general features of the experimental data which means that IQMD with
the above parameters is a reasonable transfer theoretical model for simulating
the dynamical process in intermediate energy heavy ion collisions.
\subsection{A probe of isospin dependent in-medium NN cross section
 by nuclceon emissions}
The isospin effects of the in-medium NN cross section on the physical
quantities arise from the difference between  
isospin dependent in-medium NN cross section denoted by $\sigma^{iso}$ 
in which $\sigma_{np} \geq  \sigma_{nn}$=$\sigma_{pp}$ and isospin
independent NN cross section denoted by 
$\sigma^{noiso}$ in which $\sigma_{np}$ = $\sigma_{nn}$ = $\sigma_{pp}$. 
Here $\sigma_{np}$ , $\sigma_{nn}$ and $\sigma_{pp}$ are the neutron-proton
,neutron-neutron and proton-proton cross sections respectively.
\par
Here N$_p$(N$_n$) includes all of  protons(neutrons) emitted during the nuclear reaction. 
To identify free nucleons,a phase-space coalescence method has been 
used at 200fm/c (when N$_p$(N$_n$) becomes nearly a constant) after the initial 
contact of the two nuclei. A nucleon is considered as free if it is not correlated with other 
nucleon within a spatial distance of $\Delta$r=3fm and a momentum 
distance of $\Delta$p=300MeV/c.Otherwise,it is bound in a cluster.
In addition to the nucleon emissions we also calculated all of the fragments
during the same reaction .  
Fig.2 shows the time evolutions of the N$_n$(top windows) 
and N$_p$(bottom windows) for the colliding systems $^{76}$Kr+$^{40}$Ca, 
$^{74}$Kr+$^{74}$Se,$^{76}$Kr+$^{76}$Kr and $^{74}$Se+$^{74}$Se with 
neutron-proton ratios 1.04,1.06,1.11 and 1.18 respectively at impact 
parameter b=4.0 fm and beam energy E=150 MeV/nucleon.To identify the 
isospin effects of the two-body collision on the N$_n$(N$_p$) we 
compare the cases when the isospin dependence of in-medium NN cross section is 
either turned on ($\sigma^{iso}$) or off($\sigma^{noiso}$).Turning on 
the isospin-dependent NN cross section is seen to enhance the momentum dissipation
as expected,leading to a larger number of proton(neutron) emissions.The isospin effects
of the one-body dissipation and two-body collsion on the N$_n$(N$_p$)
 are identified by using five cases
:$U_1^{sym} + \sigma^{iso}$ (solid lines), $U_0^{sym} + \sigma^{iso}$ (dot lines),
$U_2^{sym} + \sigma^{iso}$ (dashed lines) ,$U_1^{sym} + 
\sigma^{noiso}$ (dashed-dot lines), $U_2^{sym} + \sigma^{noiso}$ 
(dashed-dot-dot lines).From Fig.2 it is clear to see that the 
variations among the values of N$_p$(N$_n$) with the same 
$\sigma^{iso}$ or the same $\sigma^{noiso}$ but different symmetry 
potentials are smaller but the gaps between the N$_p$(N$_n$) with 
$\sigma^{iso}$ and  N$_p$(N$_n$) with $\sigma^{noiso}$ are larger, 
i.e,the values of N$_p$(N$_n$) depends sensitively on the 
isospin dependence of in-medium NN cross section and weakly on 
the symmetry potentials for the neutron-deficient systems.
This situation is contrary to the neutron-proton ratio of the preeqilibrium
emissions in relatively low energy ergions,where the neutron-proton ratio of the
preeqilibrium emissions depends sensitively on the symmetry potential and
weakly on the isospin dependence of the in-medium NN cross section $^{[19]}$.
With decreasing beam energy the 
role of two-body collision is reduced.In particular the symmetry potential is repulsive
for the neutrons and attractive for the protons which
tends to make more neutrons than protons unbound i.e.the role of symmetry potentail
on the nucleon emissions at low energies results in the sizeable differences
between neutron emissions and proton emissions,but the two-body collision
produces about the same probability for gaining enough energy to become unbound for
the the neutrons and protons.As a result,the isospin dependence
of in-medium NN cross section has a small affect on the neutron-proton ratio 
of the preeqilibrium emissions but the role of the symmetry potential on it
is strengthened in regions of relatively low beam energy.But with increasing beam energy 
the role of two-body collisions increases while the role of the mean 
field is reduced i.e.,in the relatively high energy region the two-body collision 
is dominant.Especially in this paper we calculated the neutron emissions and
the proton emissions not the ratio of them.Finally we can get the results as in Fig.2.
In Fig.3 is shown the impact parameter-averaged asymptotic values of N$_n$(top window)
and N$_p$(bottom window) as a function of beam energy for the systems $^{76}$Kr+$^{40}$Ca 
and $^{76}$Kr+$^{76}$Kr in the five cases considered as the same as Fig.2.From Fig.3 it is
very clear to see the same conclusion as Fig.2 in the beam energy region from about 50  to 400 MeV/nucleon.
But as the beam energy is decreased to about 50 MeV/nucleon the N$_p$(N$_n$) depends on both the
isospin dependence of in-medium NN cross section and the symmetry potential.It is worth 
mentioning that the minimum energies for remaining above property are small different for the 
systems with the variation of the neutron-proton ratios and mass asymmetries of the 
colliding systems.
\par
The above results show strongly that N$_p$(N$_n$) during the neutron-deficient nuclear reaction
can be used as a sensitive probe for extracting information on the isospin dependent
in-medium NN cross section.Here the above behaviour has been shown for the
neutron emissions and proton emissions separately,of couse,there is also the same behaviour as
a probe for the sum of them.\\\
 It is worth mentinoing that in general,the correction effect of the sequential 
 decays on the dynamical process of HIC and the number of the nucleon emissions should be 
 considered.For example M.B.Tsang et al'studies
 show that the apparent temperatures measured with double ratios of fragment isotope
 yields display fluctuations that can be attributed to the sequential decay of heavier particle
 unstable nuclei $^{[37]}$.However the number of nucleon emissions in our calculations is the
 final total number of nucleon emissions after colliding system has reached at equilibrium 
 which includes the original nucleon emissions and sequential decays leading above conclusion.
 Namely the final total number of nucleon emissions dependes sensitively on the isospin effect of 
 in-medium NN cross section and insensitively on the symmetry potential as a 
 probe for extracting information on the isospin dependent in-medium NN cross section in HIC 
 at the relatively high beam energies for neutron-deficient colliding systems.Even 
 though the dynamical mechanisms for the original nucleon emissions and the sequential decays
 are different which produces the correction effect on the total number of  
 nucleon emissions but this difference and the correction effect of sequential decays on the total 
 number of nucleon emissions do not influence our final conclusion.
\section{Summary and conclusions}
\hskip 0.3in In summary,within the IQMD we studied the isospin effects of one-body dissipation and 
two-body collision on the number of protons(neutrons) emitted during the nuclear reaction.
The calculated results show strongly that the isospin-dependent in-medium
NN cross section has a much stronger influence on N$_p$(N$_n$) but the effects of the symmetry 
potential on them are smaller for the neutron-deficent systems in the relatively high energy 
region.Studies of N$_p$(N$_n$) during the nuclear reactions are proposed to extract information
on the isospin-dependent in-medium NN cross section in HIC based on systimatic comparisons
between the theoretical simulations and experimental studies.
\par
This work was supported in part by the Major State Research
Development Project in China under Contract No. G2000077400,100 person
Project of the Chinese Academy of Sciences, the National Natural Foundation
of China under Grants No.10175080,No.10004012 and The CAS
Knowledge Innovation Project No.KJCX2-SW-N02,

\baselineskip 0.2in
\section*{Figure captions}
\begin{description}
\item[Fig. 1]
The correlation between the mean intermediate mass
fragment multiplicity $N_{imf}$ and the charged particle
multiplicity $N_{c}$. Filled (unfilled) circles represent the
experimental data [36] for the reactions $^{58}Ni
+^{58}Ni$($^{58}Fe+^{58}Fe$) at E=75 MeV/u and the solid line
(dot line) indicates the IQMD results for $^{58}Ni + ^{58}Ni$(
$^{58}Fe+^{58}Fe$).The charge number of $N_{imf}$ is taken from 3
to 18.
\item[Fig. 2]  The time evolutions of the N$_n$(top windows) 
and N$_p$(bottom windows) for the colliding systems $^{76}$Kr+$^{40}$Ca, 
$^{74}$Kr+$^{74}$Se, $^{76}$Kr+$^{76}$Kr and $^{74}$Se+$^{74}$Se 
with neutron-proton ratios 1.07,1.11,1.11 and 1.18 respectively at impact 
parameter b=4.0fm and beam energy E=150MeV/nucleon for five cases (see text).
\item[Fig. 3]
The impact parameter-averaged  
values of N$_n$(top window) and N$_p$(bottom window)as a function of the beam
energies for the systems $^{76}$Kr + $^{40}$Ca 
and $^{76}$Kr+$^{76}$Kr in the five cases as the same as Fig.2. 
\end{description}

\begin{thebibliography}{34}
\baselineskip 0.28in
\bibitem{s1} M.S. Hussein R.A.Rego and C.A. Bertulani,Phys. Rep. {\bf 201}, 279 (1993).
\bibitem{s2}B.A. Li, C.M. Ko and W. Bauer, Int. J. Mod. Phys. {\bf E7}, 147 (1998).
\bibitem{s3} B.A. Li and W. Udo Schr\"oder,Isospin Physics in Heavy-Ion Collisions 
              at Intermediate Energies (book),Nova Science Publishers, Inc (2001, New York).
\bibitem{s4}H.A.Bethe, ReV.Mod.Phys.{\bf 62},801(1990) 
\bibitem{s5}C.J.Pethick and D.G.Ravenhall,Ann.ReV.Nucl.Part.Sci.{\bf 45},429(1995)
\bibitem{s6}R. Wada et al., Phys. Rev. Lett. {\bf 58}, 1829 (1987); 
\bibitem{s7}S.J. Yennello et al, Phys. Lett. {\bf B321}, 14 (1994);
Nucl.Phys.{\bf A681},317c(2001)and referces therein. 
\bibitem{s8}R. Pak et al. Phys. Rev. Lett. {\bf 78}, 1022 (1997); 
{\bf 78}, 1026 (1997);
\bibitem{s9}G.D. Westfall, Nucl. Phys. {\bf A630}, 27c (1998);
{\bf A681}, 343c (2001);
\bibitem{s10}G.J. Kunde et al., Phys. Rev. Lett. {\bf 77}, 2897 (1996);
\bibitem{s11}M.L. Miller et al, Phys. Rev. Lett. {\bf 82}, 1399 (1999);
\bibitem{s12}H. Xu et al, Phys. Rev. Lett. {\bf 85}, 716 (2000);
M.B.Tsang et al., Phys.Rev.Lett. {\bf 86} 5023 (2001).
\bibitem{s13}W. Udo Schr\"oder et al., Nucl. Phys. {\bf A681}, 418c (2001);
and references therein. 
\bibitem{s14}L.G. Sobotka et al, Phys. Rev. C{\bf 55}, R1272 (1994);
{\bf 62}, 031603 (2000).
\bibitem{s15}F. Rami et al, Phys. Rev. Lett. {\bf 84}, 1120 (2000);
\bibitem{s16}W.P.Tan et al,MSUCL-1198,Phys.Rev.{\bf C} (2001) in press.
\bibitem{s17}M. Farine, T. Sami, B. Remaud and F. Sebille, 
Z. Phys. {\bf 339}, 363 (1991).
\bibitem{s18}H. M\"uller and B.D. Serot, Phys. Rev. C{\bf 52}, 2072 (1995).
\bibitem{s19}B.A. Li et al, Phys. Rev. Lett. {\bf 76}, 4492 (1996);
 {\bf 78}, 1644 (1997);  {\bf 85}, 4221 (2000).
\bibitem{s20}G. Kortmeyer, W. Bauer abd G.J. Kunde, 
Phys. Rev. C{\bf 55}, 2730 (1997).
\bibitem{s21}M.Colonna et al, Phys.Lett.{\bf B428}, 1(1998);
V. Baran et al., Nucl., Phys. {\bf A632}, 287 (1998); 
M.Di Toro et al. Nucl.Phys.A681,{\bf 426c} (2001) and references therein.
\bibitem{s22}J. Pan and S. Das Gupta, Phys. Rev. C{\bf 57}, 1839 (1998).
\bibitem{s23}Ph. Chomaz and F. Gulminelli, Phys. Lett. {\bf B447}, 221 (1999).
\bibitem{s24}A. Hombach et al, Eur. Phys. J. {\bf A5}, 77 (1999).
\bibitem{s25}J.Y. Liu et al., Phys.Rev.Lett.{\bf 86} 975(2001);
Phys. Rev. {\bf C63} 054612 (2001); Nucl. Phys. {\bf A687}, 475 (2001).
\bibitem{s26}D. Klakow, G. Welke and W. Bauer, Phys. Rev. C{\bf 48}, 1982 (1993).
\bibitem{s27}K.Chen, Z. Fraenkel et al., Phys. Rev.  {\bf 166}, 949 (1968).
\bibitem{s28}J. Aichelin, Phys. Rep. {\bf 202}, 233 (1991); 
Ch. Hartnack et al., Euro. J. Nucl. Phys. {\bf A1}, 151 (1998).
\bibitem{s29}B.A. Li, Nucl. Phys. {\bf A681}, 434c (2001).
\bibitem{s30}  J. Aichelin et al., Phys. Rev. Lett. {\bf 58} 1926 (1987).
\bibitem{s31}  J. Aichelin et al., Phys. Rev. {\bf C37}, 2451 (1988).
\bibitem{s32}P.G.Reinhard et al ,Computational Nuclear Physics,Vol.1,
 Springer-Verlag,Berlin,1991,pp.28-50 Danielewicz,Nucl.Phys.{\bf A673},375(2000).
\bibitem{s33}G.Alkahzov et al., Nucl. Phys. {\bf A280}, 365 (1977).
\bibitem{s34}G.F.Bertsch and S.D.Gupta, {\it Phys.Rep.} {\bf 160}(1988)1991-233.
\bibitem{s35}C.Ngo, H.Ngo and S.Leray et al., {\it Phys.Rep.} A499(1989)148.
\bibitem{s36}M. L. Miller, O. Bjarki and D. J. Magestro et al.,
    {\it Phys.Rev.Lett.}{\bf 82}(1999)1399.
\bibitem{s37}M.B.Tsang,W.G.Lynch and H.Xi ,{\it Phys.Rev.Lett.} {\bf 78}(1997)3836-3839.
\end{thebibliography}
\end{document}